\documentclass[twocolumn,preprintnumbers,amsmath,amssymb]{revtex4}
\usepackage{graphicx}

\usepackage{amsmath}
\usepackage{amssymb}
\usepackage{amsthm}
\usepackage{amsfonts}
\usepackage{enumerate}

\usepackage{centernot}
\usepackage{tikz}

\begin{document}

\title{ Majorana fermions in three dimensions and realization in critical Weyl semimetals}

\author{Xi Luo$^{1,2}$, Feng Tang$^{3,5}$, Xiangang Wan$^{3,5}$ and Yue Yu$^{4,5}$}
\affiliation {${}^1$  College of Science, University of Shanghai for Science and Technology, Shanghai 200093, PR China \\
${}^2$ CAS Key Laboratory of Theoretical Physics, Institute of
Theoretical Physics, Chinese Academy of Sciences, P.O. Box 2735,
Beijing 100190, China\\
${}^{3}$National Laboratory of Solid State Microstructures and School of
Physics, Nanjing University, Nanjing 210093, China\\
${}^{4}$State Key Laboratory of Surface Physics, Center for Field Theory and
Particle Physics, Department of Physics, Fudan University, Shanghai 200433,
China\\
${}^{5}$Collaborative Innovation Center of Advanced Microstructures, Nanjing
210093, China}

\date{\today}

\begin{abstract}
 We present two band models for free fermion with charge conjugation symmetry in three dimensions. Without time reversal symmetry (TRS), the weak pairing gapless $A$-phase is a Majorana fermion $p_x+ip_y$ wave FFLO state while the strong pairing gapped $B$-phase belongs to topologically trivial Class $D$. With TRS, there is a  Majorana fermion $B$-phase belonging to Class $DIII$ with a non-zero Hopf invariant. The  TRS $A$-phase is also a Majorana fermion  FFLO state with TRS. The surface states of the TRS $B$-phase are either a valley-momentum locked Majorana-Dirac cone or a linear-quadratic mixed cone for a specific surface.  
 The surface states of the $A$-phase on one surface are topologically nontrivial, either having  $\mathbb{Z}$ or $\mathbb{Z}_2$ invariant depending on whether the system is TRS or not. The edge states of that surface are gapless Majorana modes.  The Majorana fermion gapless FFLO states can be realized in critical Weyl semimetals (WSM) in which dual single Weyl nodes form dipoles and are nearly annihilated. The  gapped $B$-phase emerges when Weyl node dipoles are about  to be created.  The WSM TaAs-family, a type-II WSM series Mo$_x$W$_{1-x}$Te$_2$-family, possible WSM La/LuBi$_{1-x}$ Sb$_x$Te$_3$  and  topological crystalline insulators  Sn$_{1-x}$Pb$_x$(Te,Se) are  candidates to be manipulated into these critical states based on Majorana fermion models. 
  \end{abstract}


\maketitle

\noindent{\it Introductions.}  Searching for Majorana fermions in condensed matter systems has been on-going since the concept of Majorana fermion was reintroduced to the condensed matter physics community by Moore and Read \cite{MR}. It gains a wide interest in the physics community because the Majorana bound state which is a bound state of a vortex and a zero mode of the Majorana fermion carries non-abelian statistics \cite{MR,wen} and non-abelian anyon serves as a qubit of topological quantum computation \cite{kit}.  

 Majorana fermion was used in a well-know exactly soluble model a long time ago, namely, the Ising model for classical spins which has a Majorana fermion representation. Moore and Read showed that the Majorana fermion may appear in $\nu=\frac{5}2$ fractional quantum Hall effect and the non-ablian anyon introduced by them is conformally equivalent to the vortex in the Ising model \cite{MR}.  

Read and Green (RG) proved that the asymptotic limit of Moore-Read Pfaffian state of $\nu=\frac{5}2$ fractional quantum Hall effect is the Majorana fermion in a  $p_x+ip_y$ wave weak-pairing phase, the chiral $p$-wave superconductor \cite{RG}.  A direct generalization of the chiral p-wave
Dirac equations to the three-dimensions is by adding a non-relativistic kinetic energy term, and the Dirac fermion is still the Majorana one \cite{qi,wen2,wen3,ko}.  In this sense, the three-dimensional Majorana fermion exists, e.g., in a $^3$He superfluid. The Bogoliubov quasiparticles are the Fourier modes of the Majorana fermion. However, this kind of three-dimensional Majorana fermion is emerged from a two-dimensional relativistic theory in which the non-relativistic quadratic kinematic term serves as a momentum dependent mass. We call this kind of Majorana fermion RG's Majorana fermion. It is still a task to look for the relativistic Majorana fermion in 3+1 dimensions that Majorana originally proposed in 1937 \cite{Maj}, because a relativistic generalization of the RG's Bogoliubov-de Genes (BdG) equations to three dimensions does not support a 3+1 relativistic Majorana fermion.      

Besides the $p$-wave superconductor, there are many continuous persistent efforts in finding the Majorana fermion, Majorana zero mode and Majorana bound state in one- and two-dimensional condensed matter systems, both theoretically and experimentally \cite{kit2,sau,alicea,nadj,bra,sun,He}. 

 In this article, we would like to propose a  generic RG's Majorana fermion model in three dimensions, which is a two band free fermion model with a charge conjugation symmetry. If the time reversal symmetry(TRS) is broken, it is a direct generalization of the RG's BdG system. The ground state of the weak pairing gapless phase ($A$-phase) is a $p_x+ip_y$-wave Fulde-Ferrell-Larkin-Ovchinnikov (FFLO) state \cite{FF,LO}.  The strong pairing gapped phase ($B$-phase) is in Class $D$ and topologically trivial \cite{sch,ryu,kit1}. For the system with TRS, the FFLO $A$-phase is gapless and  modulated by two TRS paired finite wave vectors which label Majorana points. This is a chiral symmetric system as the FFLO state at one Majorana point is the left chiral $p$-wave's while it is a right chiral $p$-wave's at the TRS dual Majorana point. Topological classification of the gapped $B$-phase belongs to Class $DIII$ \cite{sch,ryu,kit1} and the topological invariant is the Hopf invariant, an integer $\mathbb{Z}$, which resembles the superfluid of $^3$He \cite{volobook}.  

For the TRS $B$-phase,  there are single Majorana-Dirac cones with  valley-momentum  locking on one surface while linear-quadratic mixed cones can be observed on the other orthogonal surfaces. The number of the cones are determined by the bulk Hopf number.  For the $A$-phase, there is a gapped topologically nontrivial surface while the states on other two orthogonal surfaces are trivial. The topological invariant on that nontrivial surface is either $\mathbb{Z}$ or $\mathbb{Z}_2$ depending on whether the system is TRS or not. The edge states are either chiral gapless Majorana mode or two Majorana modes with opposite chiralities.  
   
It is difficult to find a real material whose global band structure has a charge conjugation symmetry  without interaction. It may exist in the vicinity of a Majorana point in Brillouin zone in the long wave length limit which we will discuss below. It was known that a Weyl node is a Berry curvature monopole in momentum space and a Fermi arc can be seen at the surface of a Weyl semimetal (WSM) \cite{wan}.  We will show that the Majorana point may appear either  when two dual single Weyl nodes are nearly annihilated in a WSM or such a pair of Weyl points is nearly created. We call these materials critical WSM and discuss the material realizations by manipulating Weyl nodes.

\noindent{\it Models and Majorana valleys. }  We study a generic two band model whose spectrum can be expanded at a given point ${\bf k}^m=(k^m_x,k^m_y,k^m_z)$ in Brillouin zone.  The second quantized Hamiltonian is given by 
$
\hat H=\sum_{\bf q}\psi^\dag_{{\bf k}}H({\bf k}^m,{\bf q})\psi_{{\bf k}}
$ where  ${\bf q}={\bf k}-{\bf k}^m$; $\psi^T_{\bf k}=(\psi_{1\bf k},\psi_{2\bf k})$ with $\psi_{s\bf k}$ the fermion annihilation operator and the Hamiltonian matrix $H$ reads
\begin{eqnarray}
H ({\bf k}^m,{\bf q})=f_0({\bf k}^m,{\bf q})\sigma^0+ f_i({\bf k}^m,{\bf q})\sigma^i, \label{h1}
\end{eqnarray}    
 where $\sigma^i$ are Pauli matrices and $\sigma^0$ is the $2\times 2$ identity matrix; $f_\mu$ may be expanded as a power series of ${\bf q}$
 \begin{eqnarray}
 f_\mu({\bf k}^m,{\bf q})&=& f_\mu({\bf k}^m,0)+\left(\frac{\partial f_\mu}{\partial k^i}\right)_{k=k^m}q^i\nonumber\\&+&\frac{1}2\left(\frac{\partial^2 f_\mu}{\partial k^i\partial k^j}\right)_{k=k^m}q^iq^j+\cdots.
 \end{eqnarray}
 They are convergent in Brillouin zone. Let us consider a special kind of $f_\mu$ that obeys the following three conditions:

(i) All even terms in $f_0$ are zero \cite{note}. The chemical potential is then zero and the electrons are at half-filling; 
 
(ii) keep only the even terms in one of 
 $f_i$. Without loss of generality, we take $f_i=f_z$; 
 
(iii) there are only odd terms in other two $f_i=f_{x,y}$.
 
 For the reason given below, we call ${\bf k}^m$ Majorana point if $f_\mu$ satisfy these three conditions. The neighborhood in the vicinity of the Majorana point is called a Majorana valley. Notice that the number of the Majorana points does not suffer the lattice doubling for the numbers of Dirac or Weyl points \cite{NN}.   A simple example of such kind of systems is a two band tight-binding model in a cubic lattice %with $H_{lattice}({\bf k})=\sin k_x\sigma^x+\sin k_y\sigma^y+(\sum_{i=1}^3\cos k_i-2+m_z)\sigma^z+(m_z-1)\sigma^0$
 which is used to construct a general Weyl semimetal \cite{kyyang,Li}. This system is a generalized RG's Hamiltonian with a ${\bf k}$-dependent mass term \cite{RG} and then describes a $p$-wave superconductor.  It may satisfy the three conditions and the Majorana point is at $\Gamma=(0,0,0)$. Another example is Weyl metal defined in a bipartite array of lattice
planes with spinless fermions \cite{Ho,Hald,YL}. Majorana point is also at $\Gamma$.  
 In general, Eq. (\ref{h1}) shows that the Majorana point has a finite wave vector. A simplest example we will study is given by
 \begin{eqnarray}
 H_{1MP}=f_0 + v_xq_x\sigma^x+v_yq_y\sigma^y+v_z(m+q^2_z)\sigma^z, \label{h1mp}
 \end{eqnarray}
where $f_0$ is a linear term in $q_i$ and whose spectrum is
$E({\bf q})=f_0 \pm [v^2_xq_x^2+v_y^2q^2_y+(m+q^2_z)^2]^{1/2} .$  This spectrum is depicted in Fig. 1. 

\begin{figure}	
\includegraphics[width=0.45\textwidth]{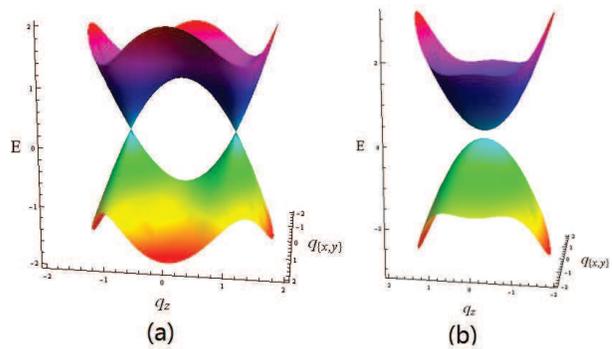}
	\caption{ (Color online) The spectra of Hamiltonian (\ref{h1mp}) with $f_0=0, v_x=v_y=v_z=1$, (a) $m=-1.0$ and (b) $m=0.1$. 
		\label{fig1}	}
\end{figure}

 The model (\ref{h1}) can  be generalized to that with multi-Majorana valleys,  e.g., if the system has TRS, we then have two Majorana points which are TRS dual. For example, Eq. (\ref{h1mp}) can be generalized to that at valleys in the vicinity of  Majorana points $\pm {\bf k}^m$ 
 \begin{eqnarray}
\hat H_{2MP}&=&\sum_{{\bf q},s=\pm}\psi^\dag_{s{\bf k}^m+{\bf q}}H_s\psi_{s{\bf k}^m+{\bf q}}+\cdots,\label{2mp}
 \end{eqnarray}
 where 
 $ 
 H_\pm=\pm(f_0 +v_xq_x\sigma^x+  v_yq_y\sigma^y)+v_z(m+q^2_z)\sigma^z,
 $ 
 and $"\cdots"$ stands for high energy terms far from the Fermi level and will be explained later after we study the discrete symmetries of the Hamiltonian.   

\noindent{\it  Charge conjugation symmetry and Majorana fermion.}  We now prove the Hamiltonian (\ref{h1}) obeying the three conditions is a Majorana fermion theory. In real space, the electron field is defined by $\psi^T({\bf r},t)=(\psi_1({\bf r},t),\psi_2({\bf r},t))^T=\sum_{\bf q}(\psi_{1\bf k},\psi_{2\bf k})^Te^{i{\bf q}\cdot{\bf r}-iEt}$. In the continuum limit, the quantum equations of motion of $\psi({\bf r},t)$ are given by
\begin{eqnarray}
i\frac{\partial\psi ({\bf r},t)}{\partial t}=H({\bf r})\psi ({\bf r},t), \label{1}
\end{eqnarray}
where 
$
H({\bf r})=f_\mu({\bf k}^m,-i\nabla)\sigma^\mu
$ .
  The charge conjugation operator ${\cal C}$ is defined by
$
 C\psi C=\psi^c={\cal C}\psi=\sigma^xK\psi=\sigma^x\psi^*
 $
 where $\psi^*=(\psi^\dag)^T$ and ${\cal C}^2=1$ \cite{supp}.
  Explicitly, 
 \begin{eqnarray}
\psi^c_1=\psi^\dag_2,~ \psi^c_{2}= \psi_1^\dag. \label{cc}
\end{eqnarray} 
  
Write Eq. (\ref{1}) as the generalized Dirac equations of RG's Dirac equations \cite{RG} 
\begin{eqnarray}
&&i\frac{\partial\psi_1}{\partial t}=(f_0+f_z) \psi_1 +(f_x-if_y)\psi_2 ,\nonumber\\
&&i\frac{\partial\psi_2}{\partial t}=(f_0-f_z) \psi_2 +(f_x+if_y)\psi_1. \label{RGD}
\end{eqnarray}
Following Read and Green \cite{RG}, $\psi^\dag_1=\psi_2$ is a necessary condition for any solution of (\ref{RGD}) with the three conditions. Substituting this to Eq. (\ref{cc}), we see that the fermion field automatically obeys Majorana condition 
\begin{eqnarray}
\psi^c=\psi~{\rm or}~\psi^\dag_2({\bf r},t)=\psi_1({\bf r},t),~\psi^\dag_{2,{\bf k}^m+\bf q}=\psi_{1,{\bf k}^m-\bf q}.   \label{mc}
\end{eqnarray}
This means that the two-band model obeying the three conditions is a Majorana fermion model.  In fact, this is a result of the charge conjugation invariance of the Hamiltonian matrix $H({\bf r})$, ${\cal C}H({\bf r}){\cal C} =H({\bf r})$.

\noindent{\it Topological properties and time reversal symmetric model.} The Hamiltonian (\ref{h1}) does not have TRS, i.e., $\Theta H\Theta^{-1}\ne H$ for $\Theta=i\sigma^y K$. This model belongs to Class $D$ as a chiral $p$-wave superconductor \cite{sch,ryu,kit1}. In three dimensions, this Class is topologically trivial. Precisely, it is an FFLO state because of the finite wave vector of the Cooper pairs \cite{FF,LO}.

We now consider $H_{2MP}$, Eq. (\ref{2mp}). When $m>0$, the spectrum of $H_{2MP}$ is fully gapped while there are four nodes if $m<0$, which resembles the $B$-phase and $A$-phase in  superfluid $^3$He, respectively \cite{volobook}.
This Hamiltonian owns two TRS dual Majorana points and it is TRS, i.e., invariant under $H_\pm\to\Theta H_\pm\Theta^{-1}$ and $\psi_{\bf k}\to \Theta \psi_{-\bf k}$.  It is also of a chiral symmetry and is charge conjugation invariant in the sense of ${\cal C}H_\pm({\bf r}){\cal C}=H_\pm({\bf r})$ with ${\cal C}=\sigma^xK$. The $"\cdots"$ in $H_{2MP}$ may break this charge conjugation symmetry but they are far from the Fermi level and can be neglected when studying the long wavelength physics.

The topological classification of $H_{2MP}$  belongs to Class $DIII$.  In three dimensions, this Class labels chiral symmetric $p$-wave topological superconductor with a topological number belonging to $\mathbb{Z}$ \cite{sch,ryu,kit1}. Here, the ground state is a chiral symmetric $p$-wave FFLO state. To define the topological invariant, we compactify the long wavelength wave vector space $\{{\bf q}\}$ into a three sphere $S^3$.  The Hamiltonian $H_{2MP}$ induces a map from $f:S^3\to S^2$, a Hopf mapping. The homotopy group $\pi_3(S^2)=\pi_3(S^3)=\mathbb{Z}$ gives a topological invariant, the Hopf invariant which describes the linking number between two curves in $\{{\bf q}\}$ \cite{wz}.  A circle $S^1$ of $S^3$ is mapped to a point on $S^2$. For example, $S^1_+$ and $S^1_-$ with reference points $( \hat q_x, \pm \hat q_y,\pm\hat q\sin\varphi_0,
\hat q\cos\varphi_0)$ where $\hat q=[1-\hat q^2_x-\hat q^2_y]^{1/2}$ are mapped to points  in the neighborhoods  of the north  and the south poles of $S^2$, respectively:
\begin{eqnarray}
f:&&S^1_\pm= \{( \hat q_x, \pm \hat q_y,  \pm \hat q\sin(\varphi+\varphi_0), \hat q\cos(\varphi+\varphi_0))\} \nonumber \\
&&\to \frac{( q_x,\pm q_y,\pm |m+q_{z}^2|)}{\sqrt{q^2_x+q^2_y+(m+q^{2}_{z})^2}},
\end{eqnarray}
where $\varphi$ runs from 0 to $2\pi$ and $q_{z}=A(\varphi_0)$ is a real function of $\varphi_0$.
 In the $B$-phase, $m>0$, two neighborhoods can connect and make a two sphere $S^2$. The Hopf invariant is +1. In the $A$-phase, $m<0$, the point  $(0,0,\pm|m+q_{z}^2|)$ is a null point if $q_{z}^2=|m|$.  This means that $f$ maps $S^3$ to a topologically trivial space. Thus, the $A$-phase is topologically trivial. The topological properties of the $A$- and $B$-phases here  resemble those in the $A$- and $B$-phase of superfluid $^3$He \cite{volobook}.
 
If there are multi-pairs of the TRS dual Majorana points, every pair contributes one to Hopf invariant and the total Hopf invariant is given by the number of the pairs of the dual Majorana points.    
 
 \begin{figure}	
\includegraphics[width=0.45\textwidth]{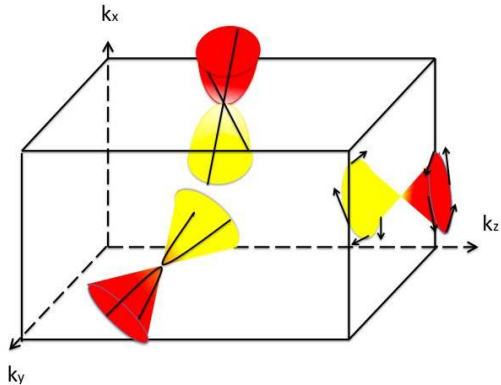}	
\caption{ (Color online) The schematics of the surface gapless states. On the surface with a constant $k_z$, [001]-plane, one can see a Majorana-Dirac cones. The arrows labels the valley-momentum locking, analogous to the spin-momentum locking in the topological insulator \cite{qi1}.  On the surface with a constant $k_x$ (or $k_y$), [100] (or [010]) plane, the surface cone is linear in the $k_y$ ($k_x$) direction while quadratic in the $k_z$-direction. 
		\label{fig3}	}
\end{figure}

\noindent{\it Surface states.}  The surface-bulk correspondence can  verify the topological properties. The surface states of the Hamiltonians (\ref{h1mp}) and (\ref{2mp}) can be easily solved. There are  no topologically protected stable gapless surface states of (\ref{h1mp}) as expected. For the TRS system, 
the surface states of the $B$-phase  on the [001] surface is described by the effective Hamiltonian $H_{[001]}\propto v_xq_x\tau^x+v_yq_y\tau^y$ where the Pauli matrices $\tau^{x,y}$ respect the valley degrees of freedom \cite{chung}. Analogous to the surface states of $\mathbb{Z}_2$ topological insulator \cite{qi1},   
there is a  Majorana-Dirac cone with valley-momentum locking (See Fig. \ref{fig3} and \cite{supp}).  On each other surface, $H_{surface}\propto v_{x,y}q_{x,y}\tau^{x,y}+v_zq_z^2\tau^z$. This gives a linear-quadratic mixed cone.  Due to TRS, these cones are stable to a perturbation with TRS. For multi-pairs of the Majorana points, there will be multi-surface cones on a surface.  Counting the number of the cones on a given surface gives the Hopf invariants of the bulk. 

There is no topological protected two-dimensional gapless states on the surface of the $A$-phase while the edge states of the surface may be topologically protected.  Consider a semi-infinite system with $x_a=0$ surface. For the Hamiltonians (\ref{h1mp}) and (\ref{2mp}), the Dirac operator is reduced to a one-dimensional one which is topologically trivial on a two-dimensional surface. On the $z=0$ surface, the $x$-$y$ plane, the system is a gapped two-dimensional FFLO state which is topologically nontrivial, either having a $\mathbb{Z}$ winding number for Hamiltonian  (\ref{h1mp}) or a $\mathbb{Z}_2$ invariant for  Hamiltonian (\ref{2mp}).  On the edges of the surfaces, there is a chiral Majorana gapless mode for the TRS breaking $A$-phase while two opposite chiral gapless modes with opposite valley indices travel on the edge of the surface of the TRS $A$-phase. The vortices on the $z=0$ surface of the $A$-phases are then the non-abelian anyons because they can trap or not a Majorana zero mode, similar to that in the $B$-phase of Kitaev honeycomb model \cite{kit}.
Constructing the qubits using these vortices will be studied elsewhere.

\noindent{\it  FFLO ground states.} We now show the ground states of the system are FFLO states by using the BdG equations. We study the Hamiltonian (\ref{h1}) and denote 
 $\psi_{\bf k}=(\psi_{1{\bf k}^m+{\bf q}},\psi_{2{\bf k}^m+{\bf q}})^T=(c_{1\bf q},c_{2\bf q})^T$ and do a unitary transformation by
\begin{eqnarray} 
\alpha_{\bf q}=u^*_{+,\bf q}c_{1\bf q}+v^*_{+,\bf q}c_{2\bf q}, ~\beta_{\bf q}= u_{-,\bf q}c^\dag_{1\bf q}+v_{-,\bf q}c^\dag_{2\bf q},\label{bog}
\end{eqnarray}
so that the Hamiltonian are diagonalized as $\hat H\propto\sum_{\bf q}E_{\bf q}(\alpha^\dag_{\bf q}\alpha_{\bf q}+\beta^\dag_{\bf q}\beta_{\bf q})$.
Then, $(u_{\pm\bf q},v_{\pm\bf q})$ are the positive and negative energy solutions of the following equations corresponding to Eq. (\ref{RGD}):
\begin{eqnarray}
&&Eu_{\bf q}=(f_{0\bf q}+f_{z\bf q})u_{\bf q}+(f_{x\bf q}-if_{y\bf q})v_{\bf q},\nonumber\\
&&Ev_{\bf q}=(f_{0\bf q}-f_{z\bf q})v_{\bf q}+(f_{x\bf q}+if_{y\bf q})u_{\bf q}. 
\end{eqnarray}
One may immediately checked that there are the Majorana relations between the positive and negative solutions
$%\begin{eqnarray}
u_{-,\bf q}=v^*_{+,-\bf q},~v_{-,\bf q}=u^*_{+,-\bf q}. 
$%\end{eqnarray}
This means that $\beta_{\bf q}=v^*_{+,-\bf q}c^\dag_{1\bf q} +u^*_{+,-\bf q}c^\dag_{2\bf q}$ is not independent of $\alpha_{\bf q}$. It turns out $\beta_{\bf q}=\alpha_{-\bf q}$ \cite{supp} . The correct Hamiltonian is $\hat H=\sum_{\bf q}E_{\bf q}\alpha^\dag_{\bf q}\alpha_{\bf q}$ with $\alpha_{\bf q}=u^*_{+,\bf q}c_{1\bf q}+v^*_{+,\bf q}c^\dag_{1,-\bf q} $.  
The ground state of the system is annihilated by $\alpha_{\bf q}$, i.e, 
\begin{eqnarray}
|FFLO\rangle\propto\prod_{\bf q}\alpha_{\bf q}|0\rangle\propto \prod_{\bf q}|u_{\bf q}+v_{\bf q}c^\dag_{1\bf q}c^\dag_{1,-\bf q})|0\rangle.
\end{eqnarray}
Notice that $c^\dag_{1\pm\bf q}=\psi^\dag_{1{\bf k}^m\pm{\bf q}}$, the ground state is a $p$-wave pairing FFLO state. 
The phase with $m<0$ is the weak pairing $A$-phase \cite{RG}. It is gapless in three dimensions. A finite center-of-mass momentum of a Cooper pairs leads to a spatial modulation of the superconductor order parameter and than the spatial alternation of the superconducting and normal conducting areas. This may be detected by NMR \cite{wri,may,kou} and heat capacity measurement \cite{lor,bey,ago}. The positive $m$ $B$-phase is the strong pairing phase  \cite{RG} which is a trivial insulator. 

For the TRS model,  each pair of Majorana points has two FFLO momenta with opposite signs. In the gapless $A$-phase, the $p$-wave FF-type state gives rise to a spatial modulated order parameter $\Delta({\bf r,q}) \sim \Delta (q_x\cos ({\bf k}^m\cdot{\bf r})+ q_y \sin ({\bf k}^m\cdot{\bf r}))$. We hope this anisotropic modulation to $q_x$ and $q_y$  can be observed in the $A$-phase. % It is interesting to see that the $A$-phase is topologically trivial while the strong pairing $B$-phase is topologically nontrivial in three dimensions.  

\noindent{\it Critical Weyl semimetals and material's realization. } The Majorana points may exist in critical WSM. 
The critical WSM is named when two dual single Weyl nodes form a dipole and are nearly annihilated in WSM. In this critical region, both Weyl valley and band degrees of freedom cannot be detected as the Weyl nodes are not independent. The physical degrees of freedom reduce to one and thus the quasiparticle and quasihole bands are related by the Majorana constraint, i.e., a particle is also its anti-particle. The ground state in the Majorana valley is an FFLO state, the $A$-phase. Opposite to the annihilation of a pair of Weyl nodes, a Majorana point can also be created when a pair of Weyl points  is about to emerge in an insulator with nearly closed gap. We  call this the $B$-phase of the critical WSM. The Hamiltonian of the critical WSM is in general given by Eq. (\ref{h1}), e.g., the spectrum in the neighborhood of the Majorana point is of the shapes given in Fig. \ref{fig1}. For a critical WSM with TRS, there at least two pairs of Weyl nodes due to TRS, which results a TRS dual pair of Majorana points.

The candidate materials of the critical WSM  have already existed  when the inversion symmetry breaking WSM with TRS was predicted and experimentally discovered \cite{weng,huang,xu,lv,yang,solu}. A pair of dual Weyl nodes in momentum space are very close in the type-I WSM, TaAs family.  The spectra of TaAs family near the bottom of the band may asymptotically described by  the Hamiltonian (\ref{h1}) and are of the shape similar to Fig. \ref{fig1}(a) in long wavelength limit,  with coordinates' redefinition. The Berry curvature dipolar structure of a pair of Weyl point was studied and its effect, quantum nonlinear Hall effect, was found \cite{Sodemann}. Recently, an {\it ab initio} study showed that the Berry curvature dipoles may exist in two WSM series, TaAs-family and type-II WSM MoTe$_2$-family \cite{zhang}. Thus, the latter is also a possible candidate of the critical WSM. 

For TaAs family, there are 12 pairs of Weyl points  \cite{weng} and they may form 12 Berry curvature dipoles. Four pairs of Weyl nodes 1 are exactly at $k_z=0$ plane, say, $(0.0,\pm 0.949, \pm 0.014)$ and $(0.0,\pm0.014, \pm0.949)$ for TaAs. If a pressure or an external magnetic field is applied, two pairs of Weyl nodes may be annihilated, say, $ (0.0,\pm 0.949, \pm 0.014)\to  (0.0,\pm 0.949, 0.0)$ and then a pair of TRS dual Majorana points appear. For Mo$_x$W$_{1-x}$Te$_2$, there are eight Weyl nodes at $(0.0,\pm k^m_y, \pm k^m_z\pm \delta^{(1,2)})$ with $\delta^{(1)}<\delta^{(2)}$ \cite{mwt}. When $\delta^{(1)}=0$, we have a pair of Majorana points.  Furthermore, the gap opens when the external pressure or the magnetic field increases. A topological non-trivial $B$-phase appears. In TaAs family, there are eight pairs of Weyl nodes 2. Manipulating those Weyl nodes may lead to a Hopf invariant 2. To ensure the particle-hole symmetry, we require the Weyl point to be as close to the Fermi surface as possible.

There were already attempts to manipulate the Weyl nodes by high-pressure \cite{hir,guo,Liang}. In Ref. \cite{Liang}, there were evidences that topological crystalline insulator Pb$_{1-x}$Sn$_x$Te becomes a WSM under pressure and Weyl nodes re-annihilate as pressure increases. 
In Ref. \cite{Ra}, the critical WSM region in TaAs  has already been reached under a high magnetic field. The annihilation of Weyl nodes and a gap opening were shown.  %We does not study this critical WSM in a high magnetic field. However, the disappearance of the chiral anomaly in the $n=0$ Landau level may easily be seen. 

Doping is also a tool to manipulate WSM. Topological insulators LaBiTe$_3$ and LuBiTe$_3$ doped by Sb were predicted  to be WSM  LaBi$_{1-x}$ Sb$_x$Te$_3$ and LuBi$_{1-x}$ Sb$_x$ Te$_3$ for $x\approx  38.5\%-41.9\%$ and $x \approx 40.5\%-45.1\%$, respectively \cite{dop}. A topologically trivial state to type-II WSM transition was found in  Mo$_x$W$_{1-x}$Te$_2$ at $x\approx7\%$ \cite{dop1}.  
The magnetically-doped (Mn, Eu, Cr, et al) Sn$_{1-x}$Pb$_x$(Te,Se) topological crystalline insulators are possible to be transformed into a TRS breaking WSM \cite{Liu}. In the critical region, one expects to find the critical WSM. The disorder may lead to the annihilation and creation of Weyl nodes and the WSM-insulator transition \cite{xie}. 

\noindent{\it The Majorana representation.}  It is worth mentioning that if we  take $f_i=f_y$ in (ii) of the three conditions, say $f_y=m+q_y^2$, the charge conjugation operator becomes ${\cal C}\psi=\psi^*$ and the necessary condition of a solution of the equations of motion becomes a familiar one 
\begin{eqnarray}
\psi^\dag_1({\bf r},t) =\psi_1({\bf r},t), ~\psi^\dag_2({\bf r},t)=\psi_2({\bf r},t), 
\end{eqnarray} 
i.e., $\psi_1$ and $\psi_2$ are "real" fermions. This is the Majorana representation. The  unitary transformation (\ref{bog}) becomes 
\begin{eqnarray} 
\alpha_{\bf q}=u_{+,\bf q}c_{1\bf q}+v_{+,\bf q}c_{2\bf q}, ~\beta_{\bf q}= u_{-,\bf q}c_{1\bf q}+v_{-,\bf q}c_{2\bf q},
\end{eqnarray}
where $u_{\pm,\bf q}$ and $v_{\pm,\bf q}$ are real  and the Majorana condition reads $c_{s,\bf q}=c^\dag_{s,-\bf q}$. The ground state  is then a three-dimensional Kitaev chiral spin liquid, which was widely studied recently \cite{kit3}. We thus predict the critical WSM as an alternative kind of Kitaev materials besides the iridates and $\alpha$-RuCl$_3$ \cite{Ir,RuCl1,RuCl2,RuCl3}.

For $f_i=f_x$ in (ii), the charge conjugation operator is ${\cal C}\psi=\sigma^z\psi^*$. The Majorana condition  becomes $\psi^\dag_s({\bf r},t)=s\psi_s({\bf r},t)$.

\noindent{\it Conclusions.} We proposed  three-dimensional  RG's Majorana fermion models whose weak pairing ground state is the $p$-wave FFLO state.  For a TRS model, the $B$-phase is topologically nontrivial with a nonzero Hopf invariant. Exotic gapless Majorana surface states were found. The number of surface cones counts the Hopf invariant. The topologically nontrivial surface states of the $A$-phase were found and  the gapless Majorana edge modes were shown. We predicted that these Majorana states can be realized in existed and expected WSM systems with external field, pressure or doping manipulations. 

\noindent{\it Acknowledgements.} This work was supported by NSFC under Grants No. 11474061 (XL, YY), 11774066 (XL, YY), 11525417 (FT, XGW), and 11374137 (FT, XGW); the Ministry of Science and Technology of China (Grant number: 2017YFA0303200, FT, XGW).

%\newpage
\begin{appendix}

\section{Charge conjugation symmetry at one Majorana valley and FFLO ground state}
The effective Hamiltonian near a Majorana valley ${\bf k}^m$ in the Brillouin zone can be written as,
\begin{eqnarray}
H ({\bf k}^m,{\bf q})=f_0({\bf k}^m,{\bf q})\sigma^0+ f_i({\bf k}^m,{\bf q})\sigma^i,\label{ah1}
\end{eqnarray} 
where the coefficients $f_\mu$ satisfies the three conditions as in the main text. Therefore, after the Peierls substitution (or minimal coupling) of the external electromagnetic field,
\begin{equation}
H({\bf k}^m,{\bf q})\rightarrow H({\bf k}^m,{\bf q}-e{\bf A})+eA_0.
\end{equation}
The equations of motion for the wave function $\psi({\bf r},t)$ becomes,
\begin{equation}
i\partial_0\psi=(H({\bf k}^m,{\bf -i\nabla}-e{\bf A})+eA_0)\psi.
\end{equation}
If there exists a charge conjugation operator ${\cal C}:\psi\longmapsto \psi^c$, such that $\psi^c$ satisfies,
\begin{equation}
i\partial_0\psi^C=(H({\bf k}^m,{\bf -i\nabla}+e{\bf A})-eA_0)\psi^c,
\end{equation}
then the Hamiltonian $H$ (\ref{ah1}) has a charge conjugation symmetry. Indeed, such an operator ${\cal C}$ can be found when all three conditions are satisfied. For example, $f_z$ only contains even terms in momentum ${\bf q}$ and the other three coefficients only contain odd terms. Then ${\cal C}=e^{i\theta}\sigma^x K$, where $K$ is the complex conjugation operator, $\theta$ is a global phase and ${\cal C}^2=1$. 

Furthermore, if the equations of motion for $\psi$ is governed by the Hamiltonian (\ref{ah1}), then $\psi^c$ satisfies the same equations of motion with the same Hamiltonian. Since the Hamiltonian (\ref{ah1}) is a two level system without degeneracy, then $\psi$ and $\psi^c$ should be the same state up to a global phase which can be absorbed in the definition of charge conjugation. Therefore $\psi$ and $\psi^c$ satisfy the Majorana condition $\psi=\psi^c$ and $\psi$ becomes a Majorana quasi-particle. The emergence of the Majorana quasi-particle is a direct 3D generalization of Read and Green's Majorana fermion in 2D $p$-wave superconductor\cite{rg2d}.

To illustrate this point further, we consider the second quantization of $\psi$. Firstly transform $\psi$ and $\psi^c$ into momentum space, i.e., 
\begin{eqnarray}
\psi&=&\int\frac{d^3 q}{(2\pi)^3}
\left(
\begin{array}{c}
c_{1{\bf q}} \\ 
c_{2{\bf q}}
\end{array} \right)
\exp{(-i{\bf q\cdot x})}\label{ap1}\\
\psi^c&=&\int\frac{d^3 p}{(2\pi)^3}\left(
\begin{array}{c}
c^\dag_{2{\bf q}} \\ 
c^\dag_{1{\bf q}}
\end{array} \right)
\exp{(i{\bf q\cdot x})}\nonumber\\
&=&\int\frac{d^3 q}{(2\pi)^3}\left(
\begin{array}{c}
c^\dag_{2{\bf -q}} \\ 
c^\dag_{1{\bf -q}}
\end{array} \right)
\exp{(-i{\bf q\cdot x})},
\end{eqnarray}
where $(c_{1\bf q},c_{2,\bf q})=(\psi_{1,{\bf k}^m+{\bf q}}, \psi_{2,{\bf k}^m+{\bf q}})$ are the Fourier components of the $\psi$ field. From the Majorana condition $\psi=\psi^c$, the Fourier components $(c_1,c_2)$  satisfy, ( ${\bf k}^m$ does not change to $-{\bf k}^m$.)
\begin{equation}
c_{1{\bf q}}=c^\dag_{2{-\bf q}}, \quad 
c_{2{\bf q}}=c^\dag_{1{-\bf q}}. \label{am1}
\end{equation}
In the language of second quantization, the field $\psi$ can be written as,
\begin{eqnarray}
\psi
=\int\frac{d^3 q}{(2\pi)^3\sqrt{2}}(\phi_{+,{\bf q}}\alpha_{\bf q}+\phi_{-,{\bf q}}\beta^\dag_{\bf q})e^{-i{\bf q\cdot x}}\label{ap2}
\end{eqnarray}
where $\phi_\pm=(u^*_\pm,v^*_\pm)^T$ are the normalized eigen functions of the Hamiltonian (\ref{ah1}) associated with the eigen energy $\pm E({\bf q})$, and $\alpha(\beta)$ is the annihilation operator of the quasi-particles(holes). If $\alpha$ and $\beta$ satisfy the anti-commutation relation with the only non-zero anti-commutator,
\begin{eqnarray}
&&\{\alpha_{\bf q},\alpha^\dag_{\bf q'}\}=(2\pi)^3\delta^{(3)}({\bf q-q}'), \quad\nonumber\\
&&\{\beta_{\bf q},\beta^\dag_{\bf q'}\}=(2\pi)^3\delta^{(3)}({\bf q-q}'),
\end{eqnarray}
then the Hamiltonian is diagonalized, 
\begin{equation}
\psi^\dag H\psi\propto\int\frac{d^3q}{(2\pi)^3}\frac{E}{2}(\alpha^\dag_{\bf q} \alpha_{\bf q}+\beta^\dag_{\bf q}\beta_{\bf q} ),
\end{equation}
where we have dropped a constant zero point energy.
Now comparing Eq. (\ref{ap1}) with Eq. (\ref{ap2}), 
\begin{eqnarray}
\alpha_{\bf q}&=&u^*_{+,{\bf q}}c_{1\bf q}+v^*_{+,{\bf q}}c_{2\bf q}\\
\beta^\dag_{\bf q}&=&u^*_{-,{\bf q}}c_{1\bf q}+v^*_{-,{\bf q}}c_{2\bf q}.
\end{eqnarray}
Because of the charge conjugation symmetry of the Hamiltonian (\ref{ah1}), $u_{-,\bf q}=v^*_{+,-\bf q},~v_{-,\bf q}=u^*_{+,-\bf q}.$, together with Eq. (\ref{am1}), one can check that $\alpha_{\bf q}=\beta_{\bf -q}$ which satisfies the Majorana condition on creation and annihilation operators, namely, the quasi-particles are Majorana fermions. And the vacuum is the FFLO vacuum, i.e.,
\begin{equation}
\forall \alpha_{\bf q},\quad \alpha_{\bf q}|FFLO\rangle=0,
\end{equation}
or equivalently,
\begin{equation}
|FFLO\rangle\propto\prod_{\bf q}(u_{+,{\bf q}}+v_{+,{\bf q}}c^\dag_{1\bf q}c^\dag_{1,-\bf q})|0\rangle.
\end{equation}

If $f_y$ in (\ref{ah1}) contains only even terms in momentum ${\bf q}$ and the other three coefficients  contain only odd terms, one can check that ${\cal C}=K$. This gives Majorana representation and the Majorana condition is $\psi^c_s({\bf r},t)=\psi^\dag_s({\bf r},t)=\psi_s({\bf r},t)$ for each  band index $s$. And  if $f_x$  contains only even terms in momentum ${\bf q}$ and the other three coefficients contain only odd terms, ${\cal C}=\sigma^zK$ and $\psi^c_s({\bf r},t)=s\psi^\dag_s({\bf r},t)=\psi_s({\bf r},t)$.

\section{Surface states of  Hamiltonian $H_{2MP}$.}

\begin{figure}	
\includegraphics[width=0.35\textwidth]{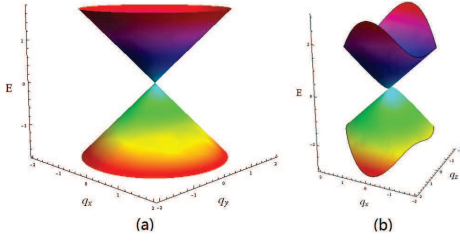}
	\caption{ (Color online) The dispersion of the surface states with $f_0=0, v_x=v_y=v_z=1$. (a) The 
linear Majorana-Dirac cone on $x$-$y$ plane. (b) The linear-quadratic mixed cone on $x$-$z$ plane.
		\label{pic}	}
\end{figure}
Let us consider the surface state of $H_{2MP}$ in the $x$-$y$ plane, i.e., $z>0$ area is the vacuum and $z<0$ is the material. In the $x$-$y$ plane $q_x$ and $q_y$ are still good quantum numbers while  $q_z\to-i\partial_z$. In order to find the gapless surface state, we  choose $q_x=q_y=0$, and seek the zero energy solution. The equations of motion for the zero energy solution read
\begin{eqnarray}
v_z(m+(-i\partial_z)^2)\sigma_z\psi_0(z)  =0
\end{eqnarray}   
The normalizable solution in the $z>0$ region reads,
\begin{equation}
\psi_0(z)=e^{-z/\lambda }(a,b)^T,
\end{equation}
where the decaying length $\lambda=1/\sqrt{m}$ and $a,b$ are normalizing constants.   Because of the charge conjugation symmetry, $a$ and $b$ are not independent of each other.  Removing the redundant antiparticle degrees of freedom and considering the valley degrees of freedom, the zero-energy states 
are given by 
\begin{eqnarray}
\Phi_1=e^{-z/\lambda}
\left(
\begin{array}{c}
c_+ \\ 
0 \\
d_- \\
0
\end{array} 
\right),\quad
\Phi_2=e^{-z/\lambda}
\left(
\begin{array}{c}
0 \\
-d^*_+\\
0\\
-c^*_-
\end{array} 
\right),
\end{eqnarray}
where $\Phi_1$ and $\Phi_2$ are related through time reversal operation and the subindex $\pm$ stands for $\pm(k^m_x+q_x,k^m_y+q_y)$. Without loss of generality, we can choose the functions $c,d$ to be real. Then the projected surface Hamiltonian reads
\begin{equation}
H^{surface}_{ij}(q_x,q_y)=\langle \Phi_i |H_{2MP}|\Phi_j\rangle=A(v_xq_x\tau^x+v_yq_y\tau^y),
\end{equation} 
where $A=-\langle c_+d_++c_-d_-\rangle$; the Pauli matrices $\tau^a$ respect for the valley degrees of freedom. Through this qualitative analysis, we discover a Majorana-Dirac cone and a valley-momentum locking phenomenon of the surface state in the $x$-$y$ plane (See Fig. \ref{pic}(a)). Analogous to the surface Dirac cone and spin-momentum locking in the topological insulator\cite{ti}, this Majorana-Dirac cone is also protected by time reversal symmetry. If there are multiple pairs of Majorana points in the bulk, each pair of them will generate a Majorana-Dirac cone on the surface. Unlike the $\mathbb{Z}_2$ classification of surface states in 3D topological insulator, the classification is   $\mathbb{Z}$ for Majorana-Dirac surface states because they belong to different valleys and the inter scattering between different valleys are difficult. This $\mathbb{Z}$ classification also shows the well known bulk-edge correspondence because the surface topological number is inherited from the bulk Hopf invariant as discussed in the main text.

The surface states on the other surfaces can be considered similarly. For example, on the $x$-$z$ plane, the effective Hamiltonian reads $A_1q_x\tau^z-A_2q_z^2\tau^x$, where $A_1,A_2$ are constants. The dispersion of surface states are shown in Fig. \ref{pic}(b).

\end{appendix}


\begin{thebibliography}{99}
\bibitem{note} When considering material realizations, we release the requirement, i.e., the leading order term is odd.
\end{thebibliography}
\end{document}